\documentclass[sigplan,screen,nonacm]{acmart}
\usepackage{bbold}
\usepackage{mathbbol}
\usepackage{ebproof}
\usepackage{setspace}
\usepackage{rotating}
\usepackage{bytefield}
\usepackage{xcolor}
\usepackage{minted}
\usepackage{mathtools}
\usepackage{circuitikz}
\usepackage{tikz}
\usepackage{todonotes}
\usepackage{mathpartir}

\setuptodonotes{inline}

\usetikzlibrary{shapes.geometric}
\usetikzlibrary{arrows.meta}

\usepackage{newunicodechar}
\newunicodechar{λ}{\ensuremath{\mathnormal\lambda}}
\newunicodechar{∀}{\ensuremath{\mathnormal\forall}}
\newunicodechar{≅}{\ensuremath{\mathnormal\cong}}
\newunicodechar{⇒}{\ensuremath{\mathnormal\Rightarrow}}
\newunicodechar{⇐}{\ensuremath{\mathnormal\Leftarrow}}
\newunicodechar{ₗ}{\ensuremath{\mathnormal l}}
\newunicodechar{ᵣ}{\ensuremath{\mathnormal r}}
\newunicodechar{≡}{\ensuremath{\mathnormal\equiv}}
\newunicodechar{⊑}{\ensuremath{\mathnormal\squbeq}}
\newunicodechar{⇛}{\ensuremath{\mathnormal\rrr}}
\newunicodechar{ᶜ}{\ensuremath{\mathnormal ^c}}

\definecolor{lightcyan}{RGB}{171, 255, 255}
\definecolor{lightyellow}{RGB}{255, 255, 171}
\definecolor{textyellow}{RGB}{235, 194, 0}
\definecolor{lightred}{RGB}{255, 171, 171}
\definecolor{lightgreen}{RGB}{171, 255, 171}
\definecolor{textgreen}{RGB}{0, 171, 0}
\definecolor{lightpurple}{RGB}{171, 171, 255}
\definecolor{lightmagenta}{RGB}{255, 171, 255}

\newcommand{\squbeq}{\sqsubseteq}
\newcommand{\vcdots}{\rotatebox{90}{$\cdots$}}
\newcommand{\rrr}{\Rrightarrow}

\newenvironment{spacedbytefield}[2][]
        {\vspace{1.5em} 
         \begin{bytefield}[#1]{#2}}
        {\end{bytefield}
         \vspace{0.2em}}

\newcommand{\bool}{\textcolor{textyellow}{\texttt{bool}}}
\newcommand{\intsmall}{\textcolor{red}{\texttt{int32}}}
\newcommand{\intbig}{\textcolor{textgreen}{\texttt{int64}}}
\newcommand{\setone}{\texttt{set1}}
\newcommand{\settwo}{\texttt{set2}}
\newcommand{\flagA}{\texttt{flagA}}
\newcommand{\flagB}{\texttt{flagB}}
\newcommand{\sizett}{\texttt{size}}
\newcommand{\payload}{\texttt{payload}}
\newcommand{\tagcyan}{\textcolor{cyan}{tag}}

\newcommand{\tycon}{\mathit{TyCon}}

\DeclareMathSymbol{\semi}{\mathrel}{bbold}{\lq\;}

\title{A Rig of Transformations}
\author{Emma Tye}
\orcid{0009-0003-2849-0443}
\affiliation{%
        \institution{University of Strathclyde}
        \city{Glasgow}
        \country{UK}
}
\email{emma.tye@strath.ac.uk}
\date{\today}
\acmDOI{10.48550/arXiv.2608.12409}
\acmISBN{}

\begin{document}

\begin{abstract}
        In high-level functional languages, the compiler often gives users little control over the runtime representation of data types. Yet how we model data structures at the program level can be different to how we want to represent them at the binary level, for efficiency or legacy reasons. Hence being able to describe data layouts and their transformations for data types is a useful and necessary part of programming, but difficult to do correctly, efficiently and ergonomically.

        We present a model of finite algebraic data types as a commutative rig (a ring without additive inverses), where the rig-equalities are modelled by isomorphisms between data. Using this approach, we can also model embedding a data type into a larger type (e.g. bit-padding) as a partial isomorphism.
\end{abstract}

\maketitle

\section{Introduction} \label{intro}

Algebraic data types (ADTs) have been widely adopted in programming languages since they can closely represent various mental models of data, are compositional and are easy to program correctly with. However, naive memory representations of ADTs can be extremely inefficient.


%
%
%
%
%

Take the following OCaml algebraic data type of settings options \texttt{convenientSet}:
\\
\begin{minipage}[t]{.22\textwidth}
\begin{minted}{ocaml}
type set1 = {
  flagA : bool;
  size : int32;
  flagB : bool;
  payload : float;
}
\end{minted}
\end{minipage}
\hfill
\begin{minipage}[t]{.22\textwidth}
\begin{minted}{ocaml}
type set2 = {
  size : int64;
  payload : float;
  flagA : bool;
}
\end{minted}
\end{minipage}

\begin{minted}{ocaml}
type convenientSet = Set1 of set1 | Set2 of set2
\end{minted}

Values of type \bool{}  use a word of memory, and \texttt{float} uses 2 words of memory \cite{madhavapeddy_real_ocaml}. The size of \intsmall{} and \intbig{} is, as expected, 32 bits and 64 bits respectively. The memory representation in bits for \setone{} and \settwo{} on a 32-bit architecture would be:

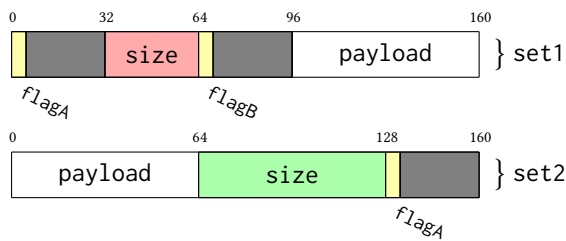
\begin{spacedbytefield}[bitwidth=0.11em, curlyshrinkage=1em]{160}
        \bitheader{0,32,64,96,160} \\
        \begin{rightwordgroup}{\setone{}}
                \bitbox{5}[bgcolor=lightyellow]{}
                & \bitbox{27}[bgcolor=gray]{}
                & \bitbox{32}[bgcolor=lightred]{\sizett{}}
                & \bitbox{5}[bgcolor=lightyellow]{}
                & \bitbox{27}[bgcolor=gray]{}
                & \bitbox{64}{\payload{}}
        \end{rightwordgroup}
        \\
        \bitbox[]{5}{\footnotesize \turnbox{-25}{\flagA}}
        & \bitbox[]{59}{}
        \bitbox[]{5}{\footnotesize \turnbox{-25}{\flagB}}
        \\
        \bitheader{0,64,128,160} \\
        \begin{rightwordgroup}{\settwo{}}
                \bitbox{64}{\payload{}}
                &\bitbox{64}[bgcolor=lightgreen]{\sizett{}}
                & \bitbox{5}[bgcolor=lightyellow]{}
                & \bitbox{27}[bgcolor=gray]{}
        \end{rightwordgroup}
        \\
        \bitbox[]{128}{}
        & \bitbox[]{5}{\footnotesize \turnbox{-25}{\flagA{}}}
\end{spacedbytefield}

The actual amount of information contained in a \bool{} is only 1 bit, so its representation contains a lot of wasted space, which we've shown in \textcolor{gray}{grey} (not to scale).

\texttt{convenientSet} is a sum type, and will therefore have an integer at the start of a value telling us what constructor we're using (either \texttt{Set1} or \texttt{Set2}). For simplicity, let's assume that the integer \textcolor{cyan}{tag} takes up a word of memory, and that all pointers are unboxed. The memory representation of \texttt{convenientSet} would be 6 words:

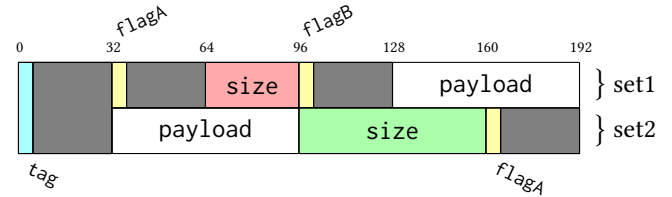
\begin{spacedbytefield}[bitwidth=0.11em, curlyshrinkage=1em]{192}
        \bitbox[]{32}{}
        & \bitbox[]{5}{\footnotesize \raisebox{0pt}[6ex][0pt]{\turnbox{25}{\flagA{}}}}
        & \bitbox[]{59}{}
        & \bitbox[]{5}{\footnotesize \raisebox{0pt}[6ex][0pt]{\turnbox{25}{\flagB{}}}}
        \\
        \bitheader{0,32,64,96,128,160,192} \\
        \begin{rightwordgroup}{set1}
                \bitbox[lrt]{5}[bgcolor=lightcyan]{}
                & \bitbox[lrt]{27}[bgcolor=gray]{}
                & \bitbox{5}[bgcolor=lightyellow]{}
                & \bitbox{27}[bgcolor=gray]{}
                & \bitbox{32}[bgcolor=lightred]{\sizett{}}
                & \bitbox{5}[bgcolor=lightyellow]{}
                & \bitbox{27}[bgcolor=gray]{}
                & \bitbox{64}{\texttt{payload}}
        \end{rightwordgroup}
        \\
        \begin{rightwordgroup}{set2}
                \bitbox[lrb]{5}[bgcolor=lightcyan]{}
                & \bitbox[lrb]{27}[bgcolor=gray]{}
                & \bitbox{64}{\texttt{payload}}
                & \bitbox{64}[bgcolor=lightgreen]{\sizett{}}
                & \bitbox{5}[bgcolor=lightyellow]{}
                & \bitbox{27}[bgcolor=gray]{}
        \end{rightwordgroup}
        \\
        \bitbox[]{5}{\footnotesize \turnbox{-25}{\texttt{tag}}}
        & \bitbox[]{155}{}
        & \bitbox[]{5}{\footnotesize \turnbox{-25}{\flagA}}
\end{spacedbytefield}

The default representation for \texttt{convenientSet} is not very efficient. Both the \bool{} and the sum \tagcyan{} only require 1 bit of information. We also need to check the tag to access the shared \sizett{} and \payload{} fields, to know their starting index and how much memory they use.

If we were to pack the \bool{}s into the first word for the sum tag, and align the \sizett{} and \payload{} fields on all constructors, we would have a more efficient representation of only 5 words. To achieve a similar result, an OCaml programmer would need to manually add extra constructors to accommodate the flags, swap the \sizett{} and \payload{} fields on \settwo{} and convert the \intsmall{} \sizett{} field in \setone{} to \intbig{}. They would then have to laboriously change all the operations using that value throughout the program. Below we show the modified OCaml definition of the data type. Each constructor name details the original sum tag and what variations of \flagA{} and \flagB{} they set (either \textbf{T}rue, \textbf{F}alse or \textbf{I}rrelevant in \settwo{}'s case, since the original definition doesn't contain a \flagB{}):

\begin{minted}{ocaml}
type set1 = { size : int64; payload : float; }
type set2 = { size : int64; payload : float; }
type efficientSet =
  (* set1 constructors *)
    Set1TT of set1 (* both flags set *)
  | Set1TF of set1 (* just flagA set *)
  | Set1FT of set1 (* just flagB set *)
  | Set1FF of set1 (* no flags set *)
  (* set2 constructors *)
  | Set2TI of set2 (* flagA set *)
  | Set2FI of set2 (* no flags set *)
;;
\end{minted}

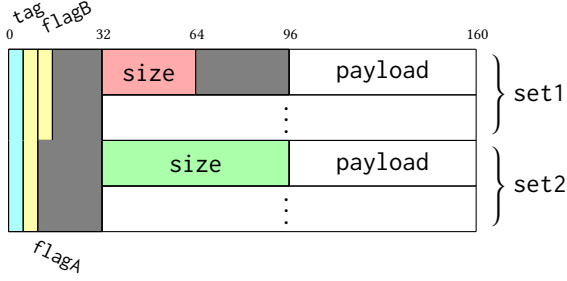
\begin{spacedbytefield}[bitwidth=0.11em, curlyshrinkage=1em]{160}
        \bitbox[]{5}{\footnotesize \raisebox{0pt}[6ex][0pt]{\turnbox{25}{\texttt{tag}}}}
        & \bitbox[]{5}{}
        & \bitbox[]{5}{\footnotesize \raisebox{0pt}[8ex][0pt]{\turnbox{25}{\flagB{}}}}
        \\
        \bitheader{0,32,64,96,160}
        \\
        \begin{rightwordgroup}{\setone{}}
                \bitbox[lt]{5}[bgcolor=lightcyan]{}
                & \bitbox[lt]{5}[bgcolor=lightyellow]{}
                & \bitbox[lt]{5}[bgcolor=lightyellow]{}
                & \bitbox[lrt]{17}[bgcolor=gray]{}
                & \bitbox{32}[bgcolor=lightred]{\sizett{}}
                & \bitbox{32}[bgcolor=gray]{}
                & \bitbox{64}{\payload{}}
                \\
                \bitbox[l]{5}[bgcolor=lightcyan]{}
                & \bitbox[l]{5}[bgcolor=lightyellow]{}
                & \bitbox[lb]{5}[bgcolor=lightyellow]{}
                & \bitbox[lr]{17}[bgcolor=gray]{}
                & \bitbox{128}{\vcdots}
        \end{rightwordgroup}
        \\
        \begin{rightwordgroup}{\settwo{}}
                \bitbox[l]{5}[bgcolor=lightcyan]{}
                & \bitbox[l]{5}[bgcolor=lightyellow]{}
                & \bitbox[lr]{22}[bgcolor=gray]{}
                & \bitbox{64}[bgcolor=lightgreen]{\sizett{}}
                & \bitbox{64}{\payload{}}
                \\
                \bitbox[lb]{5}[bgcolor=lightcyan]{}
                & \bitbox[lb]{5}[bgcolor=lightyellow]{}
                & \bitbox[lrb]{22}[bgcolor=gray]{}
                & \bitbox{128}{\vcdots}
        \end{rightwordgroup}
        \\
        \bitbox[]{5}{}
        & \bitbox[]{5}{\footnotesize \turnbox{-25}{\flagA{}}}
\end{spacedbytefield}

The labour involved in changing the \intsmall{} operations to use \intbig{} operations can be mitigated by using views \cite{wadler_views}, but it's possible the view is incorrect. For example, the above \texttt{efficientSet} definition gives no indication about how the \setone{} \intsmall{} is embedded in the \intbig{}, and different views could be used on the same field to describe left or right padding.

\section{Transformation rig} \label{rig}

We now present a lightweight way to seamlessly program using \texttt{convenientSet}, while being able to specify having an \verb|efficientSet| runtime representation, using the widely studied algebraic nature of sums and products \cite{ml_history_macqueen}. Sum and product types get their names from their set-theoretic semantics - the (finite) cardinality of types as sets of values is added under "sum" and multiplied under "product" \cite{mccarthy_computation}.

Isomorphisms between sum and product types form a proof-relevant rig, which we use to mediate between high-level and efficient runtime representations of data. Two data types that are equal under the rig axioms can be used interchangeably, and a correct-by-construction conversion algorithm is automatically derived from the proof.

\subsection{Theory} \label{rig:theory}

We introduce a language of simple data type descriptions as an algebraic signature:

\begin{align*}
        \tycon &: Set \\
        \mathbb{0} &: \tycon & \mathbb{1} &: \tycon \\
        + &: \tycon^2 \to \tycon & \times &: \tycon^2 \to \tycon \\
        \textbf{Sum}_n &: \tycon^n \to \tycon & \textbf{Prod}_n &: \tycon^n \to \tycon
\end{align*}

We add the $\textbf{Sum}$ and $\textbf{Prod}$ operators to model multiple constructors in a data type and their arguments respectively. This equips us with a flat sum type using a single tag to discriminate between constructors, and a flat product type providing random access to each field.

We also introduce a binary equivalence relation $\cong$ on $\tycon$ describing the equational theory of commutative rigs, with the flat $\textbf{Sum}_n$ and $\textbf{Prod}_n$ declared equivalent to left-nested folds over their binary counterparts:

\begin{align*}
        \mathbb{0} + A &\cong A &\TirName{id+}\\
        (A + B) + C &\cong A + (B + C) &\TirName{assoc+}\\
        A + B &\cong B + A &\TirName{comm+}\\
        \mathbb{1} \times A &\cong A &\TirName{id}\times\\
        (A \times B) \times C &\cong A \times (B \times C) &\TirName{assoc}\times\\
        A \times B &\cong B \times A &\TirName{comm}\times\\
        \mathbb{0} \times A &\cong \mathbb{0} &\TirName{absorb}\mathbb{0}\\
        A \times (B + C) &\cong (A \times B) + (A \times C) &\TirName{distr}\\
        \textbf{Sum}_n (A_1, ..., A_n) &\cong (...(A_1 + A_2) + ...) + A_n &\TirName{sum}\\
        \textbf{Prod}_n (A_1, ..., A_n) &\cong (...(A_1 \times A_2) \times ...) \times A_n &\TirName{prod}
\end{align*}

We also require $\cong$ to be a congruence with respect to $+,\times,\textbf{Sum}_n$ and $\textbf{Prod}_n$.

\subsection{Model} \label{rig:model}

We demonstrate this theory on a small language of finite primitives, tuples and disjoint unions:

\begin{align*}
        (\texttt{BaseType}s) \; b := \; &\textbf{Char} \mid \textbf{Float} \mid \textbf{Int8} \mid \textbf{Int16} \mid \textbf{Int32} \\
                                        &\mid \textbf{Int64} \mid \textbf{Bool} \\
        (\texttt{Types}s) \; A,B &:= b \mid \mathbb{0} \mid \mathbb{1} \mid A + B \mid A \times B \\
                                 &\mid \textbf{Sum}_n (A_1, ..., A_n) \mid \textbf{Prod}_n (A_1, ..., A_n)
\end{align*}

Our standard model is the above types, with $\mathbb{0}$ being the empty type, $\mathbb{1}$ being the unit type, binary sums and products and $n$-ary sums and products. We use the same syntax as the theory to make clear how to interpret the signature.

This is the free algebra of our signature over the \texttt{BaseType}s.

We model the equivalence relation $A \cong B$ as isomorphisms between $A$ and $B$. Each axiom is easily realised in this model, e.g. $\langle \texttt{id}, \texttt{id} \rangle$ modelling $\TirName{refl}$ and $\langle \texttt{swap}, \texttt{swap} \rangle$ modelling $\TirName{comm}\times$.

It's clear that we can add extra isomorphisms on our \texttt{BaseType}s. One of interest for our example is $\TirName{bool-sum}: \textbf{Bool} \cong \mathbb{1} + \mathbb{1}$.

\subsection{Binary Representation} \label{rig:rep}

Every term in the rig signature has a fixed size binary representation. Binary and $n$-ary sums have an 8-bit integer at the start tagging which injection we're using, and binary and $n$-ary products lay out their components sequentially. To simplify, we restrict the largest $\textbf{Sum}_n$ arity to be $2^8 - 1$, so we can fit the discriminator tag into 8-bits. $\mathbb{1}$ has an empty binary representation, and $\mathbb{0}$ has no binary representation.

We can statically calculate the size of any binary representation inductively by:
\begin{align*}
        size(\mathbb{0}) &= \bot \\
        size(\mathbb{1}) &= 0 \\
        size(A + B) &= 8 + \max{(size(A),size(B))} \\
        size(A \times B) &= size(A) + size(B) \\
        size(\textbf{Sum}_n (A_1, ..., A_n)) &= 8 + \max_{1 \le i \le n}{(size(A_i))} \\
        size(\textbf{Prod}_n (A_1, ..., A_n)) &= size(A_1) + ... + size(A_n)
\end{align*}

The layouts for the (non-zero) binary representations of terms are:

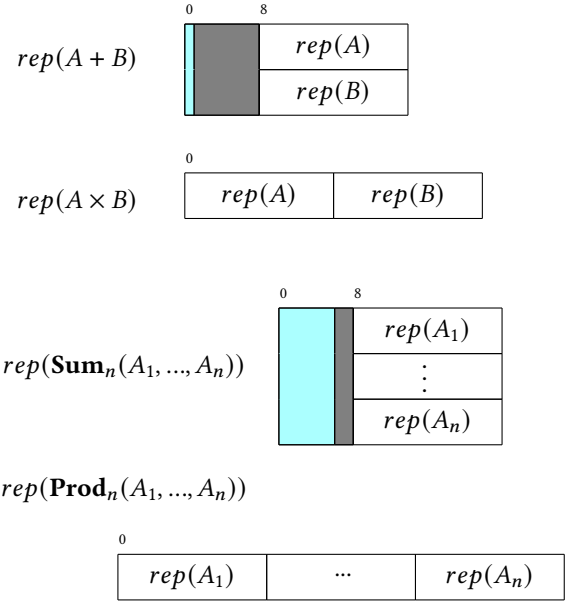
\begin{minipage}{.15\textwidth}
\[rep(A + B)\]
\end{minipage}
\begin{minipage}{.22\textwidth}
\begin{spacedbytefield}[bitwidth=0.35em]{24}
        \bitheader{0,8}
        \\
        \bitbox[lt]{1}[bgcolor=lightcyan]{} &
        \bitbox[lrt]{7}[bgcolor=gray]{} &
        \bitbox{16}{$rep(A)$} \\
        \bitbox[lb]{1}[bgcolor=lightcyan]{} &
        \bitbox[lrb]{7}[bgcolor=gray]{} &
        \bitbox{16}{$rep(B)$}
\end{spacedbytefield}
\end{minipage}

\begin{minipage}{.15\textwidth}
\[rep(A \times B)\]
\end{minipage}
\begin{minipage}{.22\textwidth}
\begin{spacedbytefield}[bitwidth=0.35em]{32}
        \bitheader{0}
        \\
        \bitbox{16}{$rep(A)$} &
        \bitbox{16}{$rep(B)$}
\end{spacedbytefield}
\end{minipage}

\begin{minipage}{.22\textwidth}
\[rep(\textbf{Sum}_n(A_1, ..., A_n))\]
\end{minipage}
\begin{minipage}{.22\textwidth}
\begin{spacedbytefield}[bitwidth=0.35em]{24}
        \bitheader{0,8}
        \\
        \bitbox[lt]{6}[bgcolor=lightcyan]{} &
        \bitbox[lrt]{2}[bgcolor=gray]{} &
        \bitbox{16}{$rep(A_1)$}
        \\
        \bitbox[l]{6}[bgcolor=lightcyan]{} &
        \bitbox[lr]{2}[bgcolor=gray]{} &
        \bitbox{16}{\vcdots}
        \\
        \bitbox[lb]{6}[bgcolor=lightcyan]{} &
        \bitbox[lrb]{2}[bgcolor=gray]{} &
        \bitbox{16}{$rep(A_n)$}
\end{spacedbytefield}
\end{minipage}

\begin{minipage}{.22\textwidth}
\[rep(\textbf{Prod}_n (A_1, ..., A_n))\]
\end{minipage}
\\
\begin{minipage}{.12\textwidth}
        \hfill
\end{minipage}
\begin{minipage}{.22\textwidth}
\begin{spacedbytefield}[bitwidth=0.35em]{32}
        \bitheader{0}
        \\
        \bitbox{16}{$rep(A_1)$} &
        \bitbox{16}{...} &
        \bitbox{16}{$rep(A_n)$}
\end{spacedbytefield}
\end{minipage}

This means that representations are not invariant under the rig equivalences - the representation for $(A + B) + C$ uses two sum-tags, whereas $\textbf{Sum}_3 (A,B,C)$ only uses one.

The size and representation of the model is given by the size and representation of its theory, along with standard representations \cite{fog_optimizing_2006} for the \texttt{BaseType}s. The size of each \texttt{BaseType} is:
\begin{align*}
        size(\textbf{Char}) &= 8 & size(\textbf{Float}) &= 32 \\
        size(\textbf{Int8}) &= 8 & size(\textbf{Int16}) &= 16 \\
        size(\textbf{Int32}) &= 32 & size(\textbf{Int64}) &= 64 \\
        size(\textbf{Bool}) &= 8
\end{align*}

The only \texttt{BaseType} with an interesting representation is \textbf{Bool}, which has an 8-bit representation, but only the first bit stores interesting information:

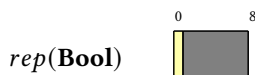
\begin{minipage}{.15\textwidth}
        \[rep(\textbf{Bool})\]
\end{minipage}
\begin{minipage}{.22\textwidth}
\begin{spacedbytefield}[bitwidth=0.35em]{32}
        \bitheader{0,8}
        \\
        \bitbox{1}[bgcolor=lightyellow]{} &
        \bitbox{7}[bgcolor=gray]{}
\end{spacedbytefield}
\end{minipage}

Our example type \texttt{convenientSet} from the introduction example would be translated into our model as:
\begin{align*}
        \textbf{Sum}_2 \bigl( &\textbf{Prod}_4 \left( \textbf{Bool}, \textbf{Int32}, \textbf{Bool}, \textbf{Float} \right), \\
                              &\textbf{Prod}_3 \left( \textbf{Float}, \textbf{Int64}, \textbf{Bool} \right) \bigr)
\end{align*}

which would then have the 112-bit binary representation:
\\
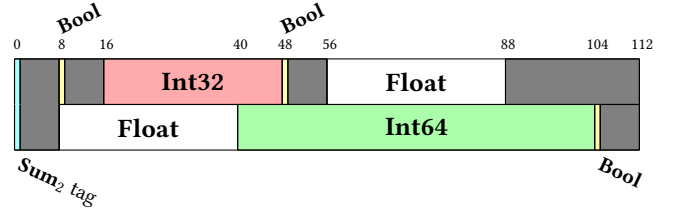
\begin{spacedbytefield}[bitwidth=0.21em]{112}
        \bitbox[]{8}{}
        & \bitbox[]{1}{\footnotesize \raisebox{0pt}[6ex][0pt]{\turnbox{25}{\textbf{Bool}}}}
        & \bitbox[]{39}{}
        & \bitbox[]{1}{\footnotesize \raisebox{0pt}[6ex][0pt]{\turnbox{25}{\textbf{Bool}}}}
        \\
        \bitheader{0,8,16,40,48,56,88,104,112}
        \\
        \bitbox[lt]{1}[bgcolor=lightcyan]{} &
        \bitbox[lrt]{7}[bgcolor=gray]{} &
        \bitbox{1}[bgcolor=lightyellow]{}
        \bitbox{7}[bgcolor=gray]{} &
        \bitbox{32}[bgcolor=lightred]{\textbf{Int32}} &
        \bitbox{1}[bgcolor=lightyellow]{} &
        \bitbox{7}[bgcolor=gray]{} &
        \bitbox{32}{\textbf{Float}} &
        \bitbox{24}[bgcolor=gray]{} 
        \\
        \bitbox[lb]{1}[bgcolor=lightcyan]{} &
        \bitbox[lrb]{7}[bgcolor=gray]{} &
        \bitbox{32}{\textbf{Float}} &
        \bitbox{64}[bgcolor=lightgreen]{\textbf{Int64}} &
        \bitbox{1}[bgcolor=lightyellow]{} &
        \bitbox{7}[bgcolor=gray]{}
        \\
        \bitbox[]{1}{\footnotesize \turnbox{-25}{$\textbf{Sum}_2$ tag}}
        & \bitbox[]{103}{}
        & \bitbox[]{1}{\footnotesize \turnbox{-25}{\textbf{Bool}}}
\end{spacedbytefield}

To transform this type closer to \texttt{efficientSet},we compose isomorphisms using transitivity of $\cong$ in the following way (omitting symmetricity, associativity and congruence rules for simplicity):
\begin{align*}
        &\textbf{Sum}_2 ( \textbf{Prod}_4 (...) \; , \; \textbf{Prod}_3 (...) ) \\ 
        \cong \; &(\textbf{Bool} \times \textbf{Int32} \times \textbf{Bool} \times \textbf{Float}) & \\
                 &+ \, (\textbf{Float} \times \textbf{Int64} \times \textbf{Bool}) &\TirName{sum}, \TirName{prod} \\
        \cong \; &((\mathbb{1} + \mathbb{1}) \times \textbf{Int32} \times (\mathbb{1} + \mathbb{1}) \times \textbf{Float}) & \\
                 &+ \, (\textbf{Float} \times \textbf{Int64} \times (\mathbb{1} + \mathbb{1})) &\TirName{bool-sum} \\
        \cong \; &((\mathbb{1} + \mathbb{1} + \mathbb{1} + \mathbb{1}) \times \textbf{Int32} \times \textbf{Float}) & \\
                 &+ \, ((\mathbb{1} + \mathbb{1}) \times \textbf{Int64} \times \textbf{Float}) &\TirName{comm}\times, \TirName{distr} \\
        \cong \; &(\textbf{Int32} \times \textbf{Float}) + (\textbf{Int32} \times \textbf{Float}) \\
                 &+ \, (\textbf{Int32} \times \textbf{Float}) + (\textbf{Int32} \times \textbf{Float}) \\
                 &+ \, (\textbf{Int64} \times \textbf{Float}) + (\textbf{Int64} \times \textbf{Float}) &\TirName{distr} \\
        \cong \; &\textbf{Sum}_6 (\textbf{Int32} \times \textbf{Float} \; , \; ... \; , \; \textbf{Int64} \times \textbf{Float}) &\TirName{sum}
\end{align*}

The $\textbf{Sum}_6 (...)$ type has a more compact 104-bit representation:

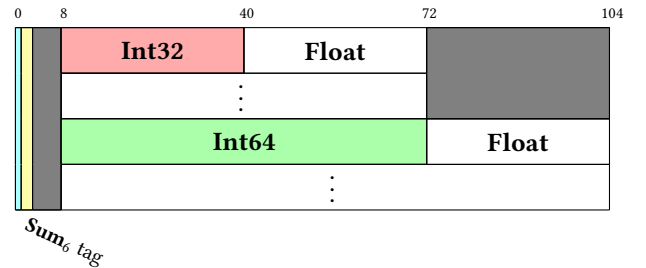
\begin{spacedbytefield}[bitwidth=0.215em]{104}
        \bitheader{0,8,40,72,104}
        \\
        \bitbox[lt]{1}[bgcolor=lightcyan]{} &
        \bitbox[lt]{2}[bgcolor=lightyellow]{} &
        \bitbox[lrt]{5}[bgcolor=gray]{} &
        \bitbox{32}[bgcolor=lightred]{\textbf{Int32}} &
        \bitbox{32}{\textbf{Float}} &
        \bitbox[lrt]{32}[bgcolor=gray]{} 
        \\
        \bitbox[l]{1}[bgcolor=lightcyan]{} &
        \bitbox[l]{2}[bgcolor=lightyellow]{} &
        \bitbox[lr]{5}[bgcolor=gray]{} &
        \bitbox{64}{\vcdots}
        \bitbox[lrb]{32}[bgcolor=gray]{} 
        \\
        \bitbox[l]{1}[bgcolor=lightcyan]{} &
        \bitbox[l]{2}[bgcolor=lightyellow]{} &
        \bitbox[lr]{5}[bgcolor=gray]{} &
        \bitbox{64}[bgcolor=lightgreen]{\textbf{Int64}} &
        \bitbox{32}{\textbf{Float}} 
        \\
        \bitbox[lb]{1}[bgcolor=lightcyan]{} &
        \bitbox[lb]{2}[bgcolor=lightyellow]{} &
        \bitbox[lrb]{5}[bgcolor=gray]{} &
        \bitbox{96}{\vcdots}
        \\
        \bitbox[]{3}{\footnotesize \turnbox{-25}{$\textbf{Sum}_6$ tag}}
\end{spacedbytefield}

\section{Partial isomorphisms} \label{pariso}

One problem with the DSL above is that we can't control where extra space should be padded for the smaller \texttt{size} field. For sum types, any extra space on smaller sized constructors is automatically added to the end of the representation. Ideally we'd like to place the extra space directly after the \textbf{Int32}, so that the \payload{} fields align for quicker access.

\subsection{Theory} \label{pariso:theory}

To enable users to express these kinds of optimisations, we add a notion of a preorder between representations, $A \squbeq A'$. Intuitively, $A \squbeq A'$ means that $A'$ is "bigger" than $A$, through padding or adding other redundant data. Then we can right-pad our \textbf{Int32} type into an \textbf{Int64}.

This geometric intuition that the right hand side is "bigger" suggests that padding should be witnessed by $A \squbeq A \times B$. However, if $B = \mathbb{0}$, then we can derive $A \squbeq A \times \mathbb{0} \cong \mathbb{0}$ through the $\TirName{absorb}\mathbb{0}$ law. If we interpret $\mathbb{0}$ as the empty type, this violates the right-hand-side is "bigger" intuition.

Instead, using preorder axioms (transitivity and reflexivity), congruence rules for the operators of our syntax and congruence over $\cong$:

\begin{align*}
        A &\cong B & A &\squbeq A' & A' &\cong B \\
          &\implies & B &\squbeq B'
\end{align*}

we can derive a consistent version of padding:
\begin{align*}
        && A &\cong \mathbb{1} \times A &\TirName{id}\times \\
        \mathbb{1} &\squbeq B \implies & &\squbeq B \times A &\TirName{cong}\times^{\squbeq}
\end{align*}

This requires the precondition that $\mathbb{1} \squbeq B$ - which can be interpreted in a suitable model as $B$ is non-empty. It's still possible to derive the inconsistency above if in the model $\mathbb{1} \squbeq \mathbb{0}$, but it's not derivable in the theory alone.

Note that this is specifically for left-padding. A different derivation using $\TirName{comm}\times$ is necessary for right-padding.

To further restrict inconsistent models, we introduce the only non-structural axiom:

\[ \mathbb{0} \squbeq A \]

A result of reflexivity of $\squbeq$ and congruence of $\cong$ over $\squbeq$ means that the relation $\cong$ is a sub-relation of $\squbeq$ ($A \cong B \implies A \squbeq B$):
\begin{align*}
        && A &\squbeq A &\TirName{refl} \\
        A &\cong B \implies & &\squbeq B &\TirName{cong}^{\cong}\squbeq
\end{align*}

So all of the Rig axioms from Section \ref{rig:theory} are also derivable in this preorder relation. We reuse the axiom names to reference the specific pre-order relation where clear, sometimes prefixing with sym- for the reverse direction.

Although it's certainly true that
\[ A \cong B \implies A \squbeq B \and B \squbeq A \]
the reverse (anti-symmetry of $\squbeq$) is not provable in the theory.

\subsection{Model} \label{pariso:model}

We choose partial isomorphisms as our standard model of $A \squbeq A'$. A partial isomorphism $A \squbeq A'$ is a pair of functions $A \to A'$ and $A' \to \mathit{Maybe}(A)$ that are inverses of each other on the successful domain of the partial function. These can be seen as embedding-projection pairs:
\begin{align*}
        f &: A \to A' \\
        g &: A' \to Maybe(A) \\
        \forall a : A, a' : A'. \\
        f(a) = a' &\iff g(a') = \texttt{just } a
\end{align*}

Proving this is a preorder (reflexive and transitive) is straightforward, and the congruence rules are uninteresting. The non-structural axiom $\mathbb{0} \squbeq A$ follows from absurdity in the forwards direction ($f$), and the backwards function ($g$) always fails.

In this model, the forward function of $\mathbb{1} \squbeq A$ picks out a specific element of $A$, with the backwards function failing on all other values. For the non-empty types of our model, we provide these instances, chosing 0 for all numeric types, \textbf{True} for \textbf{Bool}, \textbf{a} for \textbf{Char}, the first element of any product and the first injection of any sum.

Now we can construct a partial isomorphism from \textbf{Int32} into $\textbf{Int32} \times \textbf{Int32}$, padding on the right:
\begin{align*}
        && \textbf{Int32} &\cong \mathbb{1} \times \textbf{Int32} &\TirName{id}\times \\
        && &\cong \textbf{Int32} \times \mathbb{1}  &\TirName{comm}\times \\
        \mathbb{1} &\squbeq \textbf{Int32} \therefore & &\squbeq \textbf{Int32} \times \textbf{Int32} &\TirName{cong}\times^{\squbeq}
\end{align*}

\subsection{Binary Representation} \label{pariso:rep}

We now have a DSL of the rig axioms as embedding-projection pairs to describe optimisations on ADTs. Providing a relation $A \squbeq A'$ describes how to represent a type $A$ as $A'$, which provides encoding/decoding functions that are correct-by-construction.

We have the right-padded partial isomorphism $\textbf{Int32} \squbeq \textbf{Int32} \times \textbf{Int32}$, hence we can construct a partial isomorphism from \texttt{convenientSet} into \texttt{efficientSet} using the isomorphism from Section \ref{rig:rep} and congruence rules:
\begin{align*}
        &\textbf{Sum}_2 \bigl( \textbf{Prod}_4 \left( \textbf{Bool}, \textbf{Int32}, \textbf{Bool}, \textbf{Float} \right), \\
        & \hspace{2.8em} \textbf{Prod}_3 \left( \textbf{Float}, \textbf{Int64}, \textbf{Bool} \right) \bigr) \\
        \cong \; &\textbf{Sum}_6 (\textbf{Int32} \times \textbf{Float} \; , \; ... \; , \; \textbf{Int64} \times \textbf{Float}) \\
        \squbeq \; & \textbf{Sum}_6 (\textbf{Int32} \times \textbf{Int32} \times \textbf{Float} \; , \; ... \; , \; \textbf{Int64} \times \textbf{Float})
\end{align*}

This gives us the target bit representation:

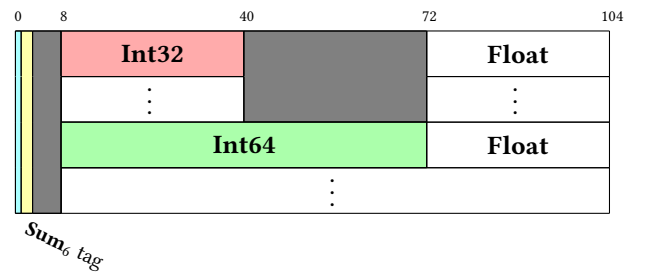
\begin{spacedbytefield}[bitwidth=0.215em]{104}
        \bitheader{0,8,40,72,104}
        \\
        \bitbox[lt]{1}[bgcolor=lightcyan]{} &
        \bitbox[lt]{2}[bgcolor=lightyellow]{} &
        \bitbox[lrt]{5}[bgcolor=gray]{} &
        \bitbox{32}[bgcolor=lightred]{\textbf{Int32}} &
        \bitbox[lrt]{32}[bgcolor=gray]{} &
        \bitbox{32}{\textbf{Float}} &
        \\
        \bitbox[l]{1}[bgcolor=lightcyan]{} &
        \bitbox[l]{2}[bgcolor=lightyellow]{} &
        \bitbox[lr]{5}[bgcolor=gray]{} &
        \bitbox{32}{\vcdots} &
        \bitbox[lrb]{32}[bgcolor=gray]{} &
        \bitbox{32}{\vcdots}
        \\
        \bitbox[l]{1}[bgcolor=lightcyan]{} &
        \bitbox[l]{2}[bgcolor=lightyellow]{} &
        \bitbox[lr]{5}[bgcolor=gray]{} &
        \bitbox{64}[bgcolor=lightgreen]{\textbf{Int64}} &
        \bitbox{32}{\textbf{Float}} 
        \\
        \bitbox[lb]{1}[bgcolor=lightcyan]{} &
        \bitbox[lb]{2}[bgcolor=lightyellow]{} &
        \bitbox[lrb]{5}[bgcolor=gray]{} &
        \bitbox{96}{\vcdots}
        \\
        \bitbox[]{3}{\footnotesize \turnbox{-25}{$\textbf{Sum}_6$ tag}}
\end{spacedbytefield}

\subsection{Using partial-isomorphisms} \label{pariso:using}

The intent behind using these partial-isomorphisms is to be able to annotate a data-type declaration with a target representation and a partial-isomorphism between the two. Some example syntax could be:

\begin{minted}[escapeinside=||,mathescape=true]{Haskell}
type eset1 = |$\textbf{Prod}_2$|(|$\textbf{Int32}$| x |$\textbf{Int32}$|, |$\textbf{Float}$|)
type eset1 = |$\textbf{Prod}_2$|(|$\textbf{Int64}$|, |$\textbf{Float}$|)
type efficientSet = 
  |$\textbf{Sum}_6$|(eset1, eset1, eset1, eset1, 
        eset2, eset2)

type set1 = |$\textbf{Prod}_4$|(|$\textbf{Bool}$|, |$\textbf{Int32}$|, |$\textbf{Bool}$|, |$\textbf{Float}$|)
type set2 = |$\textbf{Prod}_3$|(|$\textbf{Int64}$|, |$\textbf{Float}$|, |$\textbf{Bool}$|)

{# @rep: efficientSet, partial-isomorphism ... #}
type convenientSet = |$\textbf{Sum}_2$|(set1, set2)
\end{minted}
where the partial-isomorphism for \texttt{@rep} is the one described in the previous section.

In our model, partial-isomorphisms are embedding-projection pairs, and can be viewed as printers and parsers. A program using \texttt{convenientSet} would be transformed into a program on \texttt{efficientSet} by parsing the representation, applying the program, and then printing it back into the representation.

More work would need to be done to be able to translate a program on \texttt{convenientSet} into a program purely on \texttt{efficientSet} - see the Future Work Section \ref{future_work}.

\section{Meta-programs} \label{metaprograms}

Although possible for the user to programmer to write out the entire transformation for their specific data type in the above manner, it isn't erganomic or extensible. We would like to be able to describe general transformation templates without having to repeat the derivation for every possible term.

We define meta-programs to acheive this. A meta-program $Prog$ is a dependent function on terms of the algebra to a dependent pair of a new term and a partial-isomorphism between them. The latter we call a $Rep$.

\begin{align*}
        Rep(A) &:= \Sigma_{B : TyCon} \; A \squbeq B \\
        Prog &:= \Pi_{A : TyCon} \; Rep(A)
\end{align*}

Meta-programs can be composed by regular function composition and transitivity of partial-isomorphisms. In a suitable dependently-typed langauge, this would be defined as:
\begin{minipage}[t]{\textwidth}
\vspace{0.2em}
\begin{minted}[escapeinside=||,mathescape=true]{Haskell}
|$\circ$| : Prog -> Prog -> Prog
(p' |$\circ$| p) A = 
  let (B  {- : $TyCon$ -}, pariso  {- : A $\squbeq$ B -}) 
      = p A
      (C  {- : $TyCon$ -}, pariso' {- : B $\squbeq$ C -}) 
      = p' B
  in  (C, trans|$\squbeq$| (pariso, pariso'))
\end{minted}

\end{minipage}

\vspace{1em}

Application of a meta-program directly to a $Rep$ can be defined in a similar way. We use $\$$ to denote applying a meta-program to a $Rep$ and $\$\texttt{>}$ to denote backwards application of a $Rep$ to a $Prog$.

These meta-programs can be defined inductively on terms of the Rig signature via the induction principle:

\begin{minted}[escapeinside=!!,mathescape=true]{Haskell}
      -- Match on $\mathbb{0}$
ind :    Rep !$\mathbb{0}$!
      -- Match on $\mathbb{1}$
      -> Rep !$\mathbb{1}$!
      -- Match on +
      -> ((A : !$TyCon$!) -> (B : !$TyCon$!)
          -> Rep (A + B))
      -- Match on x
      -> ((A : !$TyCon$!) -> (B : !$TyCon$!)
          -> Rep (A x B))
      -- Match on $\textbf{Sum}_n$
      -> ((A1 : !$TyCon$!) -> ... -> (An : !$TyCon$!)
          -> Rep !$\textbf{Sum}_n$!(A1 ... An))
      -- Match on $\textbf{Prod}_n$
      -> ((A1 : !$TyCon$!) -> ... -> (An : !$TyCon$!)
          -> Rep !$\textbf{Prod}_n$!(A1 ... An))
      -- Describes a unique program
      -> Prog
\end{minted}

Using this induction principle, we define a custom Haskell-like match statement \texttt{repMatch} which scrutinises the given $Rep$ and uses $\$$ to chain the appropriate match clause:

\begin{minted}[escapeinside=!!,mathescape=true]{Haskell}
repMatch :    Rep A
           -- Match clauses
           -> ...
           -> Rep A
repMatch rep clauses = ind clauses $ rep
\end{minted}

We use Haskell-like syntax for match statements where you can rearrange the match clauses by specifying the structure you're matching on, allow default statements and allow matching directly on arguments to a $Prog$.

The congruence rules on our binary and $n$-ary operators also allow us to apply our operators on $Rep$s. We give the example for $+$ below:

\vspace{0.5em}

\begin{minted}[escapeinside=||,mathescape=true]{Haskell}
|$+^P$| : Rep A -> Rep B -> Rep (A + B)
  (A', pariso {- : A $\squbeq$ A' -})
|$+^P$| (B', pariso' {- : B $\squbeq$ B' -}) =
  (A' + B', cong+ (pariso, pariso')
\end{minted}

\vspace{0.5em}

The resulting partial-isomorphism is $A + B \squbeq A' + B'$. The rest of the operators have a similar definition on $Rep$s.

We can also transform the Rig axioms (as partial-isomorphisms) into meta-programs. These just apply the given single axiom to the input if appropriate, otherwise just apply $\TirName{refl}$. We give the example for $\TirName{distr}$ below:

\vspace{0.5em}

\begin{minted}[escapeinside=||,mathescape=true]{Haskell}
|$\TirName{distr}^P$| : Prog
|$\TirName{distr}^P$| (A + (B x C)) = (A x B + A x C, |$\TirName{distr}$|)
|$\TirName{distr}^P$| A = (A, |$\TirName{refl}$|)
\end{minted}

\vspace{0.5em}

We annotate axiom programs with $\_^{ -1}$ to denote meta-programs that apply their axiom symmetrically, so $\TirName{distr}^{P,-1}$ would match on terms of the form $A \times B + A \times C$ and give the representation $A + (B \times C)$.

Note that for the inverse programs of $\TirName{sum}$ and $\TirName{prod}$ we would need to supply the size of the target \textbf{Sum} and \textbf{prod}. However, the maximal size can be calculated programmatically on a given term by counting the occurrences of sub-terms appearing on the left-hand side of a $+$ or $\times$. We use $\TirName{sum}^{ -1}$ and $\TirName{prod}^{ -1}$ to denote the meta-program using the maximal size.

We can also map $Prog$s over $TyCon$s by repeatedly applying the given $Prog$ to every sub-term and using the appropriate operator combinators to combine the resulting representations. We provide pre- and post- versions of map, which apply the $Prog$ either before or after mapping over sub-terms:

\begin{minted}[escapeinside=!!,mathescape]{Haskell}
postMap : Prog -> Prog
postMap p (A + B) =
  -- Apply p after mapping over sub-terms
  p (postMap p A !$+^P$! postMap p B)
-- Similar for other binary/n-ary operators
...
-- For nullary operators (i.e. $\mathbb{0}$ and $\mathbb{1}$) 
-- and other BaseTypes, just apply p
postMap p A = p A

preMap : Prog -> Prog
preMap p A =
  -- First apply p
  repMatch p A with
  | B + C ->
    -- Then map over sub-terms
    preMap p B !$+^P$! preMap p C
  -- Similar for other binary/n-ary operators
  | ...
  -- For nullary operators (i.e. $\mathbb{0}$ and $\mathbb{1}$) 
  -- and other BaseTypes, return
  | B -> (B, !$\TirName{refl}$!)
\end{minted}

A common optimisation \cite{enum_size_rust} is to flatten nested binary and $n$-ary sums into a single \textbf{Sum}. Using the DSL defined above, pattern matching on Rig terms and recursion, we can construct a meta-program that describes this flattening optimisation for every possible type:

\begin{minted}[escapeinside=!!,mathescape=true]{Haskell}
flattenSums : Prog
flattenSums =
  -- Lastly, collapse all +s into a single $\textbf{Sum}$
    preMap !$\TirName{sum}^{ -1}$!
  -- Then, re-associate all +s to the left
  !$\circ$! postMap !$\TirName{assoc}+^{ -1}$!
  -- First, expand every $\textbf{Sum}_n$ sub-term into +s
  !$\circ$! postMap !$\TirName{sum}$!
\end{minted}

\section{Related Work} \label{related_work}

Other work on allowing finer-grained control over the runtime representation of algebraic data types includes \texttt{RIBBIT} \cite{baudon_bit-stealing_2023} and OxCaml \cite{oxcamlbox}. Our work differs from these by providing a DSL of composable primitive isomorphisms, allowing a library of possible optimisations for generic data types. There is also prior work investigating links between rigs and invertible programming \cite{Choudhury_rig_groupoid}, as well as modelling finite sets using the rig axioms \cite{elgueta_finset_rig}.

\section{Future Work} \label{future_work}

Currently the partial-isomorphisms only describe parsers and printers from a representation into a convenient type. We plan to use these partial-isomorphisms to compile programs on convenient types to programs on efficient types.

We also plan to investigate the link between our theory of partial isomorphisms to describe binary representations and rig categories \cite{Choudhury_rig_groupoid}. 

Other optimisations, such as fast lookup of aligned fields in sum types, could be added to the theory.

We are currently working with simply typed ADTs, which could be expressed using a single dependent pair $\Sigma$ type\cite{levitation_chapman}, using it for both the sum and product constructors. This would allow us to work with dependently-typed ADTs and allow correct representations of data with size components, for example.

We can also try to make our transformations themselves more efficient by normalisation. For instance, the $\TirName{id}\times$ and $\TirName{sym-id}\times$ in the following transformation cancel each other:
\begin{align*}
        A \times B &\cong (\mathbb{1} \times A) \times B &\TirName{id}\times \\
                              &\cong B \times (\mathbb{1} \times A) &\TirName{comm}\times \\
                              &\cong B \times A &\TirName{sym-id}\times
\end{align*}
making this transformation equivalent to
\begin{align*}
        A \times B &\cong B \times A &\TirName{comm}\times
\end{align*}

\bibliographystyle{ACM-Reference-Format}
\bibliography{references}

\end{document}